\documentclass[aps,prb,superscriptaddress,showpacs,amsmath,twocolumn]{revtex4}

\usepackage{amsmath}
\usepackage{amssymb}
\usepackage{graphicx}

\newcommand{\inside}[2]{\chi_{#1,#2}}
\newcommand{\outside}[3]{h_{#1,#2}^{(#3)}}
\newcommand{\kso}{k_\textrm{so}}
\newcommand{\Scat}[2]{S^{(#1)}_{#2}}

\begin{document}

\title{Two-dimensional electron scattering in regions of nonuniform spin-orbit coupling}

\author{Andr\'as P\'alyi}
\affiliation{Department of Physics of Complex Systems,
E\"otv\"os University,
H-1117 Budapest, P\'azm\'any P\'eter s\'et\'any 1/A, Hungary}

\author{Csaba P\'eterfalvi}
\affiliation{Department of Physics of Complex Systems,
E\"otv\"os University,
H-1117 Budapest, P\'azm\'any P\'eter s\'et\'any 1/A, Hungary}

\author{J\'ozsef Cserti}
\affiliation{Department of Physics of Complex Systems,
E\"otv\"os University,
H-1117 Budapest, P\'azm\'any P\'eter s\'et\'any 1/A, Hungary}

\date{\today}

\begin{abstract}
We present a theoretical study of elastic spin-dependent
electron scattering caused by a nonuniform 
Rashba spin-orbit coupling strength. 
Using the spin-generalized method of partial waves the
scattering amplitude is exactly derived for the case
of a circular shape of scattering region.
We found that the polarization of the scattered waves are strongly anisotropic
functions of the scattering angle.
This feature can be utilized to design a good all-electric spin-polarizer. 
General properties of the scattering process are also investigated 
in the high and low energy limits. 

\end{abstract}


\maketitle


The presence of the spin-orbit interaction (SOI) destroys 
the spin-rotational symmetry, therefore the properties of the
scattering get influenced by the spin state of the incident 
particle~\cite{Mott:book}.
The SOI may lead to the asymmetry of the differential scattering cross
section (skew-scattering), and it may affect the polarization vector
of the incident beam~\cite{Mott:book,Bransden:book,Bellentine:book}.
In nuclear physics, this property has been utilized to generate spin-polarized
neutrons from an unpolarized beam by scattering it on a zero-spin
nucleus\cite{Schwinger:cikk}.
In low-dimensional semiconductors, significant spin-splitting in
the absence of a magnetic field is observed, which is mostly attributed 
to the SOI of the Rashba type arising from the structural inversion
asymmetry of the hosting 
heterostructure\cite{Rashba:cikk_Winkler:book}.
Promising spin transistor application has been proposed exploiting
the tunability of the strength of the Rashba coupling by an
external electrostatic field\cite{Datta-Das:cikk}, and initiated an
intensive research in the field of spintronics\cite{spintronics-review:cikkek}.
To generate spin-polarized electron beams, which is a fundamental problem in
spintronics, several mechanisms have been 
proposed\cite{spin_pol-proposals:cikkek}. 

Recently, all-electrical (without externally applied 
magnetic fields) spin-polarizer devices have been
suggested\cite{nonuniform:cikkek} utilizing the properly designed 
spatial modulation of the Rashba SOI strength $\alpha$.  
The spatial variation of the Rashba strength, which is proportional to
the magnitude of the electric field applied perpendicular to the 
two-dimensional electron gas (2DES) systems, can be achieved 
by small biased electrodes on the top of the heterostructure. 

In this work we show that the polarization of the elastically scattered
wave caused by a nonuniform Rashba strength 
becomes strongly anisotropic, ie depends on the angle 
(called scattering angle) between the directions of the incident 
electron beam and that of the scattering wave. 
We also demonstrate that (i) the differential scattering cross section 
has a skew-scattering feature~\cite{Mott:book,Bransden:book,Bellentine:book} 
for polarized incoming electron beams, 
(ii) using experimentally relevant material parameters\cite{alpha-values:cikk}
an almost full polarization of unpolarized incident electron beams 
can be observed in a narrow window of scattering angles. 
Moreover, our analytical calculations allows us to derive universal 
properties for the scattering amplitude and the polarization 
${\bf P}^{\rm sc}$ in the high and low energy limits. 

\begin{figure}[hbt]
\includegraphics[scale=0.2]{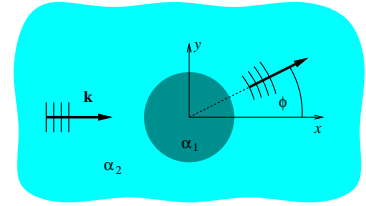}
\caption{(Color online) The plane wave of an incident electron  
with wave vector ${\bf k}$ gets scattered by a scattering region
defined by a nonuniform Rashba coupling strength $\alpha({\bf r})$ 
(see the text), and its portion propagates along the
direction given by the scattering angle $\varphi$.
\label{geo:fig}}
\end{figure}

To this end, we consider a system in which the Rashba SOI 
strength $\alpha({\bf r})$ varies on the plane of the 2DES as 
$\alpha({\bf r}) = \alpha_1 \Theta(a-|{\bf r}|) 
+ \alpha_2 \Theta(|{\bf r}|-a)$, 
where $a$ is the radius of the scattering center, 
$\Theta$ is the Heaviside function and ${\bf r}=(x,y)$ defines the
position on the plane (see Fig.~\ref{geo:fig}). 
The Hamiltonian of the system in the one-band effective-mass
approximation is given by
\begin{equation}
\mathcal{H} = \frac{{\bf p}^2}{2m^*} + 
\frac{\alpha({\bf r})}{2\hbar} (\sigma_x p_y - \sigma_y p_x)+
(\sigma_x p_y - \sigma_y p_x)\frac{\alpha({\bf r})}{2\hbar},
\label{Hamilton_def:eq}
\end{equation}
where ${\bf p}= (p_x,p_y)$ is the momentum operator, 
$m^*$ is the effective mass of the electron, $\sigma_x, \sigma_y$ are 
the Pauli matrices. 
Note that this Hamiltonian is symmetrized to make it Hermitian. 
For simplicity, we set $\alpha_2 = 0$, and $\alpha_1$ is a constant
value throughout this paper 
(our analysis can easily be extended to the case $\alpha_2 \ne 0$).
Recently, similar scattering problems have been studied with uniform 
$\alpha({\bf r})$ and an electrostatic potential varying 
in the plane of the 2DES~\cite{Walls_Heller:cikk,Yeh:cikk}.  

The spin-density matrix\cite{Bransden:book,Bellentine:book} 
is proved to be useful for treating 
the coupled spin-charge quantum transport in spintronic 
devices\cite{Nikolic-1:cikk}. 
Employing this formalism allows us to derive explicit
formulas for the differential and the total scattering cross section, 
and the polarization ${\bf P}^{\rm sc}$ of the scattered wave 
in terms of the polarization ${\bf P}^{\rm inc}$ of the incident 
electron beam.
 
As can be seen below, the physics of the scattering of electrons 
shown in Fig.~\ref{geo:fig} is fundamentally different 
from the Mott scattering originating from the spin-dependent scattering 
potential~\cite{Mott:book,Schwinger:cikk,Bransden:book,Bellentine:book}
and the spin-dependent two-dimensional electron scattering from
quantum dots and antidots studied in Ref.~\onlinecite{Voskoboynikov-1:cikk}. 
In these scattering problems the polarization ${\bf P}^{\rm sc}$ of 
the scattered wave is always perpendicular to the scattering plane for
unpolarized (${\bf P}^{\rm inc} =0$) incident particles of spin $1/2$,
while in our scattering problem this is not the case in general. 

First, we briefly outline the spin-density matrix approach of  
our spin-dependent scattering problem.
The incident plane wave with wave number ${\bf k}$ has the form
$\psi^{\rm inc} ({\bf r}) = e^{i {\bf k r}}\, | \gamma \rangle$ ,
where $|\gamma\rangle$ denotes the spin state.
In two dimensions the form of the scattered wave asymptotically 
far from the scattering center is
\begin{equation}
\psi^{\rm sc} (r,\varphi) \sim 
\frac{e^{ikr}}{\sqrt{r}} \,  {\bf f}(\varphi)\, |\gamma\rangle,
\label{f_def:eq}
\end{equation}
where ${\bf f}(\varphi)$ is the scattering amplitude ($2 \times 2$
matrix in spin space), 
and depends on the scattering angle $\varphi$ and ${\bf k}$. 
The scattering amplitude ${\bf f}(\varphi)$ can be expanded in terms
of the unit matrix $\sigma_0$ and the vector   
$\mbox{\boldmath $\sigma$}= (\sigma_1,\sigma_2,\sigma_3)$ formed from
the three Pauli matrices (for convenience, we use 
$\sigma_1 \equiv \sigma_x, \sigma_2 \equiv  \sigma_y, 
\sigma_3 \equiv \sigma_z$ notations 
for the three Pauli matrices):
\begin{equation}
\label{pauli-decomp:eq}
{\bf f}(\varphi) = \sum\limits_{k = 0}^{3} u_k(\varphi) \, \sigma_k
= u_0(\varphi) \sigma_0 + {\bf u}(\varphi) \cdot \mbox{\boldmath $\sigma$},
\end{equation}
where $ u_0$ and ${\bf u} =(u_1,u_2,u_3)$ can only be obtained 
by solving the Schr\"odinger equation for the scattering problem.

The spin-density matrix 
$\rho^{\rm inc} = \frac{1}{2}\,(\sigma_0 
+ \mbox{\boldmath $\sigma$} \cdot {\bf P}^{\rm inc})$ 
of the incident wave for a given polarization 
${\bf P}^{\rm inc} = \langle  \mbox{\boldmath $\sigma$} \rangle_{\rm inc} 
= \textrm{Tr}\, (\rho^{\rm inc} \, \mbox{\boldmath $\sigma$})$ 
is related to the spin-density matrix $\rho^{\rm sc}$ of the scattered 
wave as~\cite{Bransden:book} 
\begin{equation}
\rho^{\rm sc}  = 
\frac{{\bf f} {\rho}^{\rm inc} {\bf f}^\dag }
{\textrm{Tr}\left({\bf f} \rho^{\rm inc} {\bf f}^\dag  \right)}, 
\label{rho_scat:eq}
\end{equation}
where $\textrm{Tr}$ denotes the trace in the spin states. 
Then using (\ref{pauli-decomp:eq}) the differential scattering cross section 
reads as 
\begin{subequations}
\begin{eqnarray}
\hspace{-6mm}
\frac{d\sigma}{d\varphi}  &=& 
\textrm{Tr}\left({\bf f} \rho^{\rm inc} {\bf f}^\dag  \right) = 
c  + {\bf v}\cdot {\bf P}^{\rm inc},  
\mbox{ where} \\
\hspace{-6mm}
c &=& \sum\limits_{k=0}^{3} |u_k|^2 \,\, \mbox{ and} \,\,\,\,  
{\bf v} = 2 {\rm Re} (u_0^* {\bf u}) - i({\bf u} \times {\bf u}^*).   
\end{eqnarray}
\label{diff_sc_cross:eq}%
\end{subequations}%
Here ${\rm Re} (\cdot)$ denotes the real part of the argument, and 
the star stands for the complex conjugation. 
Note that, in general, ${\bf u} \times {\bf u}^*$ is not zero but
it is always a purely imaginary vector. 

Similarly, we found for the polarization vector ${\bf P}^{\rm sc}$ of 
the scattered beam 
\begin{subequations}
\label{sc_pol:eq}
\begin{eqnarray}
{\bf P}^{\rm sc} &=& \langle \mbox{\boldmath $\sigma$} \rangle_{\rm sc} 
= \textrm{Tr}\, (\rho^{\rm sc} \, \mbox{\boldmath $\sigma$})
= \frac{{\bf w} + \mbox{\boldmath $\mathcal{M}$} {\bf P}^{\rm inc}}
{c + {\bf v}\cdot {\bf P}^{\rm inc}},
\end{eqnarray}
where ${\bf w} = 2 {\rm Re} (u_0^* {\bf u}) + i({\bf u} \times {\bf u}^*)$ 
and the components of the matrix $\mbox{\boldmath $\mathcal{M}$}$: 
\begin{eqnarray}
\mathcal{M}_{ij} &=& 
\left(|u_0|^2 - |{\bf u}|^2\right) \delta_{ij} +
2\, {\rm Re}\, (u^*_i u_j)  \nonumber \\
&+& 2 \sum\limits_{k=1}^{3} \varepsilon_{ijk} \, {\rm Im}\, (u_0^* u_k),
\end{eqnarray}%
\end{subequations}%
with $i,j= 1,2,3$, and 
$\delta_{ij}$ and $\varepsilon_{ijk}$ denote the Kronecker delta 
and the Levi-Civita symbol, respectively.
Here ${\rm Im} (\cdot)$ stands for the imaginary part of the argument. 
The components of $\mbox{\boldmath $\mathcal{M}$}$ are real numbers.

The spin-dependent form of the optical theorem can be derived by 
considering the scattering time-evolution of Gaussian wave packets, 
and we obtain:
\begin{eqnarray}
\sigma_{\rm tot} 
&=& \sqrt{\frac{8\pi}{k}}  \, 
{\rm Im}  \left\{e^{-i\frac{\pi}{4}}  \, 
\left[u_0(0) + {\bf u}(0)\cdot {\bf P}^{\rm inc}\right] \right\}, 
\label{optic_theo:eq}
\end{eqnarray}
where $\sigma_{\rm tot}$ is the total scattering cross section. 

As can be seen, all physical quantities are expressed in terms of the 
coefficients $u_k (\varphi)$ which define the scattering amplitude 
${\bf f}(\varphi)$ in Eq.~(\ref{pauli-decomp:eq}). 
To calculate these unknown coefficients $u_k (\varphi)$ for our
scattering problem we apply the method of partial waves, similarly as
in Ref.~\onlinecite{Walls_Heller:cikk}. 
Choosing the spin quantization axis along the $z$ axis, the
eigenspinors (the eigenvectors of the Pauli matrix $\sigma_z$) 
are $\gamma_\sigma$, where $\gamma_+ = {(1,0)}^T$ for
$\sigma = +1$, and $\gamma_- = {(0,1)}^T$ for $\sigma = -1$ 
($T$ stands for the transposed of vectors). 
Hereafter, we write $\pm$ for the spin quantum number $\sigma = \pm 1$.
Since the Hamiltonian $\mathcal{H}$ in Eq.~(\ref{Hamilton_def:eq}) 
commutes with the total angular momentum operator
$J_z =-i\hbar \partial_\varphi + \frac{\hbar}{2}\,\sigma_z$, 
any partial wave, which is a solution of the Schr\"odinger equation
can be labeled by the quantum number $j \in \mathbb{J}$ 
and the spin quantum number $\sigma$ of the incident electron.  
Here $\mathbb{J} = 
\{\cdots, -\frac{3}{2},-\frac{1}{2},\frac{1}{2},\frac{3}{2},\cdots \}$. 
We chose the direction of the propagation of the incoming plane wave 
along the $x$ direction.  
Then, in polar coordinates, the incoming plane wave 
with spin quantum number $\sigma$ and energy $E$ can be expanded 
in terms of partial waves~\cite{Abramowitz-Stegun:konyv}: 
\begin{equation}
 \phi_\sigma  ({\bf r}) = e^{i k x} \gamma_\sigma = 
\frac{1}{2\sqrt{-\sigma}} 
 \sum\limits_{j \in \mathbb{J}} i^{j+1/2} 
  \left[
    \outside{j}{\sigma}{1}({\bf r})  +
    \outside{j}{\sigma}{2}({\bf r}) 
  \right],
\end{equation}
where
$\outside{j}{\sigma}{1,2}({\bf r})  =
H_{j-\sigma/2}^{(1,2)}(kr) e^{i(j-\sigma/2)\varphi}
\gamma_\sigma , 
$
are the outgoing (superscript 1) and incoming (superscript 2) waves, 
and $k = |{\bf k}|=\sqrt{2{m^*} E}/\hbar$ is the magnitude of the wave 
vector.  
Here $H_{m}^{(1,2)}(z)$ are the 1st and 2nd kind of 
Hankel functions of order $m$. 

First, we consider the individual partial waves. 
The partial waves outside the scattering region ($r>a$) have the form 
$\psi^{(\textrm{N})}_{j,\sigma} = 
 \outside{j}{\sigma}{2}  +
 \Scat{j}{\sigma,\sigma}  \outside{j}{\sigma}{1}  +
 \Scat{j}{-\sigma,\sigma}  \outside{j}{-\sigma}{1}, 
$
while inside the scattering region ($r<a$) they can be written as 
$\psi^{(\textrm{R})}_{j,\sigma} =
A^{(j)}_{+,\sigma}  \inside{j}{+}  +
A^{(j)}_{-,\sigma}  \inside{j}{-}, 
$
where
\begin{equation}
\inside{j}{\tau} ({\bf r}) =
\left(\begin{array}{c} \tau J_{j-1/2}(q_\tau r)e^{-i\varphi/2} \\[1ex]
J_{j+1/2}(q_\tau r) e^{i\varphi/2} 
\end{array}
\right)\, e^{ij\varphi},
\end{equation}
and $J_m(x)$ is the Bessel function, 
$q_\tau = \sqrt{k^2+ k_\textrm{so}^2}-\tau\kso$,  
$k_\textrm{so} =\alpha_1 m^* /\hbar^2$ and $\tau = \pm 1$ is
the spin branch index~\cite{circular_wave:eq}.  
The coefficients $A^{(j)}_{\pm,\pm} $ and $\Scat{j}{\pm,\pm}$  
can be calculated from the boundary conditions~\cite{Uli-boundary:cikk}: 
\begin{subequations}
\begin{eqnarray}
\psi^{(\textrm{N})}_{j,\sigma} \,
\rule[-1.6ex]{.2pt}{4ex} \;\raisebox{-1.5ex}{$\scriptstyle r = a$} 
&=& \psi^{(\textrm{R})}_{j,\sigma} \,
\rule[-1.6ex]{.2pt}{4ex} \;\raisebox{-1.5ex}{$\scriptstyle r = a$},\\[2ex]
\partial_r \psi^{(\textrm{N})}_{j,\sigma} \,
\rule[-1.6ex]{.2pt}{4ex} \;\raisebox{-1.5ex}{$\scriptstyle r = a$} 
&=&
(\partial_r - i\kso\sigma_\varphi) \, \psi^{(\textrm{R})}_{j,\sigma} \,
\rule[-1.6ex]{.2pt}{4ex} \;\raisebox{-1.5ex}{$\scriptstyle r = a$}, 
\end{eqnarray}
\label{boundary_cond:eq}%
\end{subequations}%
valid for all $j \in \mathbb{J}$ and $\sigma=\pm 1$, 
and $\sigma_\varphi = -\sin \varphi \, \sigma_x + \cos \varphi
\,\sigma_y$.
For a given $j$ these equations involving Bessel and Hankel functions 
result in eight linear inhomogeneous equations for the
eight unknown coefficients $A^{(j)}_{\pm,\pm} $ and
$\Scat{j}{\pm,\pm}$. 
The explicit forms of the equations are independent of the angle $\varphi$.    
These coefficients can easily be calculated numerically.
Note that the following exact relations hold 
$\Scat{j}{\sigma,\sigma} = \Scat{-j}{-\sigma,-\sigma}$ and
$\Scat{j}{-\sigma,\sigma} = \Scat{j}{\sigma,-\sigma} = 
\Scat{-j}{-\sigma,\sigma} = \Scat{-j}{\sigma,-\sigma}$ 
for all $j \in  \mathbb{J}$ and $\sigma = \pm 1$. 
To proceed further, we assume that the coefficients $\Scat{j}{\pm,\pm}$  
are known from numerical calculations. 

Outside the scattering region the complete wave function describing 
the scattering of the plane wave 
can be decomposed as  
$\psi^{(\textrm{N})}_\sigma = \sum\limits_{j\in \mathbb{J}}
\psi^{(\textrm{N})}_{j,\sigma} = 
\phi_\sigma + \psi^{({\rm sc})}_\sigma$, 
where $\psi^{({\rm sc})}_\sigma$ is the scattered wave. 
It is easy to show that $\psi^{(\textrm{N})}_\sigma$ and 
$\psi^{(\textrm{R})}_\sigma =  \sum\limits_{j\in \mathbb{J}}
\psi^{(\textrm{R})}_{j,\sigma}$ satisfy 
the boundary conditions~(\ref{boundary_cond:eq}). 
Using  the Hankel's asymptotic expansions~\cite{Abramowitz-Stegun:konyv} 
we have 
$\outside{j}{\sigma}{1}({\bf r}) \sim
\sqrt{\frac{2}{i\pi k r}}\, 
e^{i\left(kr - \left(j-\sigma/2\right)\frac{\pi}{2}\right)} \, 
e^{i(j-\sigma/2)\varphi} \, \gamma_\sigma$ 
valid for $r \gg a $, and then the asymptotic form of the scattered waves 
$\psi^{({\rm sc})}_\sigma$ can be calculated. 
Finally, Eq.~(\ref{f_def:eq}) yields 
$\langle \gamma_\sigma | \psi^{({\rm sc})}_{\sigma^\prime}
\rangle = \frac{e^{ikr}}{\sqrt{r}} f_{\sigma,\sigma^\prime}$, 
and using Eq.~(\ref{pauli-decomp:eq}) we obtain  
\begin{subequations}
\label{u_k-res:eq}
\begin{eqnarray}
 u_0(\varphi) & \!\!\!\!\! =& \!\!\!\!\!\!\! \sum\limits_{j \in \mathbb{J}^+} 
\!\! B_j  \cos  \! \left( \! j-\frac{1}{2} \! \right) \varphi 
+ C_j \cos \! \left( \! j+\frac{1}{2} \! \right)\varphi, \\
u_1(\varphi) &=& 2 \sin\left(\varphi/2\right) 
\sum\limits_{j \in \mathbb{J}^+} D_j \cos(j\varphi), \\
u_2(\varphi) &=& -2 \cos\left(\varphi/2\right) 
\sum\limits_{j \in \mathbb{J}^+} D_j \cos(j\varphi), \\
u_3(\varphi) &\!\!\!\!\!\! =& \!\!\!\!\! i \!\! 
\sum\limits_{j \in \mathbb{J}^+} 
\!\! B_j \sin \!\! \left( \! j-\frac{1}{2} \! \right)\varphi 
+ C_j \sin \!\! \left( \! j+\frac{1}{2} \! \right)\varphi, 
\end{eqnarray}%
\end{subequations}%
where we introduced the notations
$B_j = (\Scat{j}{+,+}-1)$, $C_j = (\Scat{j}{-,-}-1)$, $D_j = \Scat{j}{-,+}$
and $\mathbb{J}^+ = \{\frac{1}{2},\frac{3}{2},\cdots \} $. 

\begin{figure}[hbt]
\includegraphics[scale=0.47]{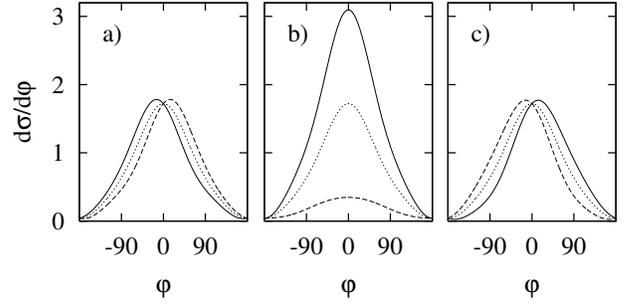}
\caption{Differential scattering cross sections 
(in units of $a$) as functions of the scattering angle $\varphi$ 
(in units of degrees) for spin polarizations 
${\bf P}^{\rm inc} = 0, {(\pm 1,0,0)}^T$ (a), 
${\bf P}^{\rm inc} = 0, {(0,\pm 1,0)}^T$ (b) and 
${\bf P}^{\rm inc} = 0, {(0,0,\pm 1)}^T$ (c)
with dotted, solid and dashed lines, respectively in each figures.  
The parameters are $ka=1$ and $\kso a=1$. 
\label{skew_cross:fig}}
\end{figure}

In numerical calculations the two dimensionless parameters
characterizing the scattering process are $ka$ and $\kso a$. 
Figure~\ref{skew_cross:fig} shows the asymmetric (skew-scattering) feature
of the differential scattering cross section calculated 
from (\ref{diff_sc_cross:eq}) for different spin polarizations 
${\bf P}^{\rm inc}$ of the incident electron beam. 
The scattering cross sections at a given scattering angle $\varphi$ 
are different for spin up and spin down polarization of an incident beam, 
ie for ${\bf P}^{\rm inc}$ and $-{\bf P}^{\rm inc}$.
From Eq.~(\ref{u_k-res:eq}) it follows that 
${\bf u}(\varphi=0) \sim  {(0,1,0)}^T$, 
therefore the optical theorem (\ref{optic_theo:eq}) implies that 
the polarization dependence of the total scattering cross section 
(the areas under the curves in Fig.~\ref{skew_cross:fig}) takes the form 
$\sigma_{\rm tot}({\bf P}^{\rm inc}) =
c_1  + c_2 P^{\rm inc}_y$, where 
$P^{\rm inc}_y $ is the $y$ component of the polarization vector
of the incident wave, and 
$c_1$ and $c_2$ depend only on $ka$ and $\kso a$.  
Thus, in Fig.~\ref{skew_cross:fig}a and c, and for 
${\bf P}^{\rm inc} = 0$ the total scattering cross sections are the same. 

\begin{figure}[hbt]
\includegraphics[scale=0.47]{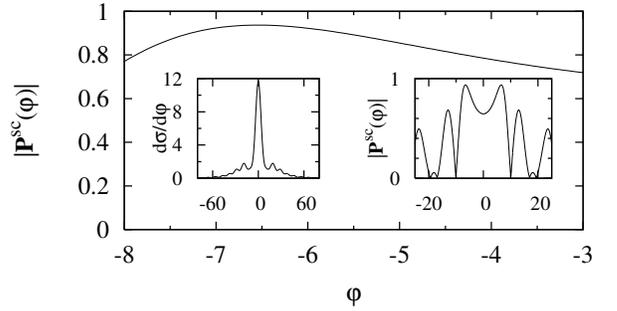} 
\caption{ 
The magnitude of polarization $|{\bf P}^{\rm sc}|$ (main panel) and 
the differential scattering cross section $d\sigma/d\varphi$ (left inset) 
as functions of the scattering angle $\varphi$ (in units of degrees) 
for unpolarized (${\bf P}^{\rm inc}=0$) incident waves.
The large scale $\varphi$ (in units of degrees) 
dependence of $|{\bf P}^{\rm sc}|$ is shown in the right inset.
The parameters are $ka=20$ and $\kso a=4$.  
\label{pol-angle:fig}}
\end{figure}

The magnitude of the polarization $|{\bf P}^{\rm sc}|$ of the
scattered waves obtained from Eq.~(\ref{sc_pol:eq}) and 
the differential scattering cross section are 
plotted in Fig.~\ref{pol-angle:fig} for unpolarized (${\bf P}^{\rm inc}=0$) 
incident electron beam using experimentally relevant 
parameters\cite{alpha-values:cikk}.  
As can be seen from the figure, 
the differential cross section is rather high and 
the scattered beam is almost fully polarized for a narrow window of angles.
This result suggests an effective tool for spin-filtering without
using magnetic field. 

We also studied the high energy limit of the scattering.
This is the case, when $ka \gg 1$, ie the Fermi wave length of the
electron is much smaller than $a$.  
Then one can apply the first Born approximation and it can be shown 
that the scattering amplitude ${\bf f} \sim \sigma_y$. 
Thus, from (\ref{pauli-decomp:eq}) we have $u_0 = 0$ and 
${\bf u} \sim {(0,1,0)}^T$, while from (\ref{sc_pol:eq}) ${\bf w}=0$. 
Therefore, for unpolarized incident electron beam 
(${\bf P}^{\rm inc}=0$) the polarization of the scattered waves is
negligible in the high energy limit.  
However, a finite spin polarization can arise even in the first Born
approximation if an additional
electrostatic potential is present beside the spin-orbit interaction, 
eg, in an imaging experiment~\cite{Walls_Heller:cikk}. 
In this case, $u_0$ and $u_2$ are finite and it yields a finite value 
of ${\bf P}^{\rm sc}$ for unpolarized incident beams. 

In the opposite limit, ie for low energy limit ($ka \ll 1$) it is
also possible to derive an analytical result for the scattering 
amplitude and the polarization  ${\bf P}^{\rm sc}$ of 
the scattered waves.  
Keeping only the first order terms in $k/\kso$ of $\Scat{j}{\pm,\pm}$ 
resulting from the boundary equations (\ref{boundary_cond:eq}) 
and that of $u_k(\varphi)$ in Eq.~(\ref{u_k-res:eq}), it yields 
\begin{equation}
{\bf P}^{\rm sc}(\varphi) \approx  
2\, \frac{k}{\kso}\, {( -\sin \varphi,1+\cos \varphi,0)}^T,  
\end{equation}
valid for $ka \ll 1$ and $k \ll \kso$ and for 
unpolarized incident electron beams. 
We found an excellent agreement between this result and that obtained 
from (\ref{sc_pol:eq}) with numerically exact calculations. 
Similarly, it can be shown that the differential scattering cross
section is approximately isotropic (independent of the scattering 
angle $\varphi$) in the low energy limit.
 
In conclusions, we have shown that for a circular shape of region 
with non-zero Rashba coupling strength, the scattering properties are 
strongly anisotropic, and the scattered wave is well polarized 
in a narrow window of the scattering directions. 
Such a nonuniform  SOI can be utilized to produce spin-polarized electrons 
in an all-electric realization of spintronic devices.  

We are grateful to I. Adagideli, B. K. Nikoli\'c, K. Richter 
and A. Csord\'as for useful discussions.
This work is supported in part by E.~C.\ Contract No.~MRTN-CT-2003-504574, 
and the Hungarian Science Foundation OTKA T034832. 

\end{document}